\documentclass[10pt]{iopart}
\usepackage[utf8]{inputenc}
\usepackage{iopams}
\usepackage{graphicx}
\usepackage{cite}
\usepackage[pdfstartview=FitH, pdfpagemode=None, pdfpagelayout=OneColumn, colorlinks=true, linkcolor=blue, citecolor = blue]{hyperref}
\newcommand{\text}[1]{\mathrm{#1}}

\begin{document}

\title[]{Generation of plasma electron halo by charged particle beam in low density plasma}
  \author{A.A. Gorn}
\address{Budker Institute of Nuclear Physics, Novosibirsk, 630090, Russia}
\address{Novosibirsk State University, Novosibirsk, 630090, Russia}
\ead{A.A.Gorn@inp.nsk.su}
\author{K.V. Lotov}
\address{Budker Institute of Nuclear Physics, Novosibirsk, 630090, Russia}
\address{Novosibirsk State University, Novosibirsk, 630090, Russia}
\ead{K.V.Lotov@inp.nsk.su}

\vspace{10pt}
\begin{indented}
\item[]\today
\end{indented}

\begin{abstract}
Breaking of a plasma wave driven by a long beam of charged particles results in electron jets escaping from the plasma column and forming an electron halo. For plasma densities less than or of the order of the beam density, this process is well described by a semi-analytical model, which agrees with simulations and allows to calculate the position of wavebreaking points and determine the regions around the plasma column occupied by the halo.
\end{abstract}

\vspace{2pc}

\ioptwocol
 \section{Introduction}
\begin{figure*}[tb]
        \includegraphics{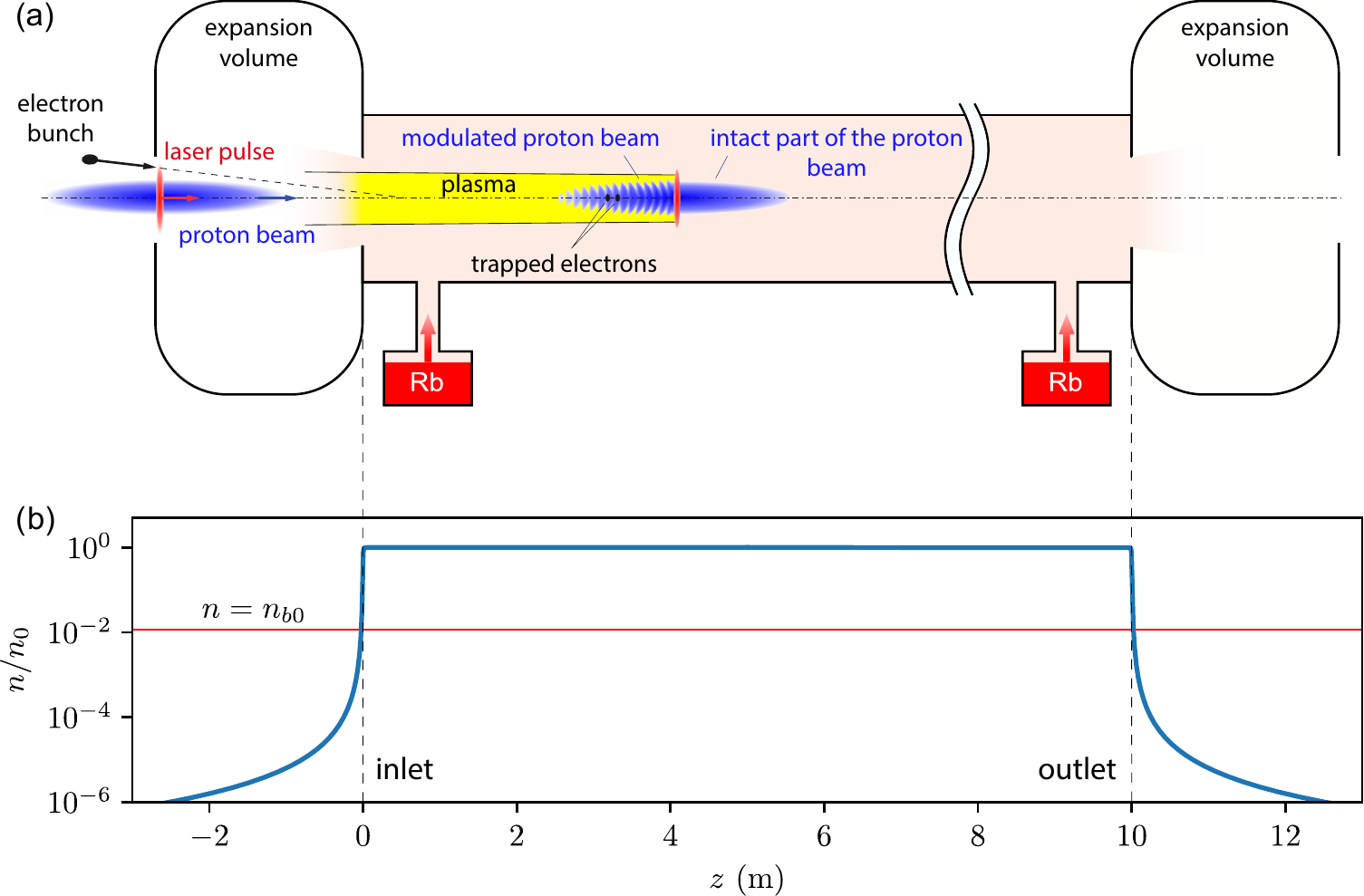}
        \caption{(a) The AWAKE schematic layout and (b) the plasma density profile $n(z)$ along the plasma cell. The red line corresponds to the maximum density of the proton beam $n_{b0} = 4\times10^{12}\,\text{cm}^{-3}$.}\label{fig1}
    \end{figure*}

    Proton-driven plasma wakefield acceleration (proton-driven PWFA) is an actively developing novel method of accelerating light charged particles \cite{NatPhys5-363,PPCF56-084013,RAST9-85}. Recently, the AWAKE project at CERN \cite{NIMA-829-3,NIMA-829-76,PPCF60-014046} demonstrated the seeded self-modulation (SSM) of a 400\,GeV proton beam \cite{PRL122-054802,PRL122-054801} and acceleration of 19\,MeV electrons to 2\,GeV in a 10\,m plasma \cite{Nat.561-363}. The plasma in the experiment is created by a short laser pulse, which co-propagates with the proton beam and single ionizes the rubidium vapor in a long gas cell (figure~\ref{fig1}(a)).
    The gas cell has an important technical feature \cite{JPD51-025203}: its ends are open and attached to the expansion volumes, so the gas can flow freely from the main vacuum chamber to the outside though the inlet and outlet orifices. Rubidium sources vaporize the liquid metal and provide a constant vapor density almost throughout the cell. This design minimizes the transition region from vacuum to the nominal density of the rubidium vapor near the both orifices in the plasma cell. However, plasma density ramps are still present and may cause issues related to the injection of the accelerated electron beam. 
    
    The proton beam generates a significant defocusing wakefield in a plasma of varying density, which destroy the electron beam injected along the axis \cite{PoP21-123116,PoP25-063108}. For this reason, it was decided to inject the accelerated beam into the plasma at a small angle \cite{NIMA-829-3}. According to simulations \cite{PoP25-063108, PRL112-194801, PPCF63-055002} confirmed by measurements \cite{PPCF62-125023,PRAB24-011301}, the interaction of the proton beam with a radially bounded plasma also leads to the ejection of plasma electrons out of the plasma boundary. The electrons gain a substantial radial momentum, taking energy from the breaking plasma wave, and escape from the plasma. This effect is closely related to the phase mixing and trajectory crossing in a cold plasma with a radial density gradient \cite{PR113-383, PoF14-1402, PRL62-2269, PoP25-022102}. The electron jets cause charge separation and create a radial electric field $E_r$ and an azimuthal magnetic field $B_\phi$ around the plasma. The radial force $F_r = -e(E_r - B_\phi)$, where $e$ is the elementary charge, can degrade the quality of the accelerated electron beam already at the injection stage. 
    
    The plasma density $n$ near the orifices of the AWAKE plasma cell has a ``ramped'' longitudinal profile (figure~\ref{fig1}(b)). It decreases towards the expansion volumes from the nominal density $n_0 = 7\times10^{14}\,\text{cm}^{-3}$ to zero according to the power law \cite{JPD51-025203}. The time between the passage of the ionizing laser pulse and the wavebreaking depends on the plasma density and hence on the longitudinal coordinate $z$. To avoid electron beam degradation outside the plasma, the beam must be injected with a limited delay after the laser pulse, that is, at $\xi = z-ct > \xi_{wb}(n(z))$, where $\xi_{wb}$ is the co-moving coordinate of the first plasma electron trajectories intersection measured from the pulse that we will call the wavebreaking point, and $c$ is the speed of light. Therefore, it is important to know how quickly an axisymmetric plasma wave breaks at different beam-plasma density ratios. In this paper, we study the formation of the plasma electron jets due to wave breaking, assuming that the wave is excited by a long, half-cut Gaussian ultrarelativistic proton beam. We develop a semi-analytical theory in section~\ref{s2} and compare it with numerical simulation results in section~\ref{s3}. We show that for the beam parameters of interest for AWAKE, the wavebreaking point can be found in the framework of a simple electrostatic model that allows analytical approximations and a fast numerical solution.

 \section{Trajectory crossing}
 \label{s2}

\begin{figure}[tb]
        \includegraphics[width=\columnwidth]{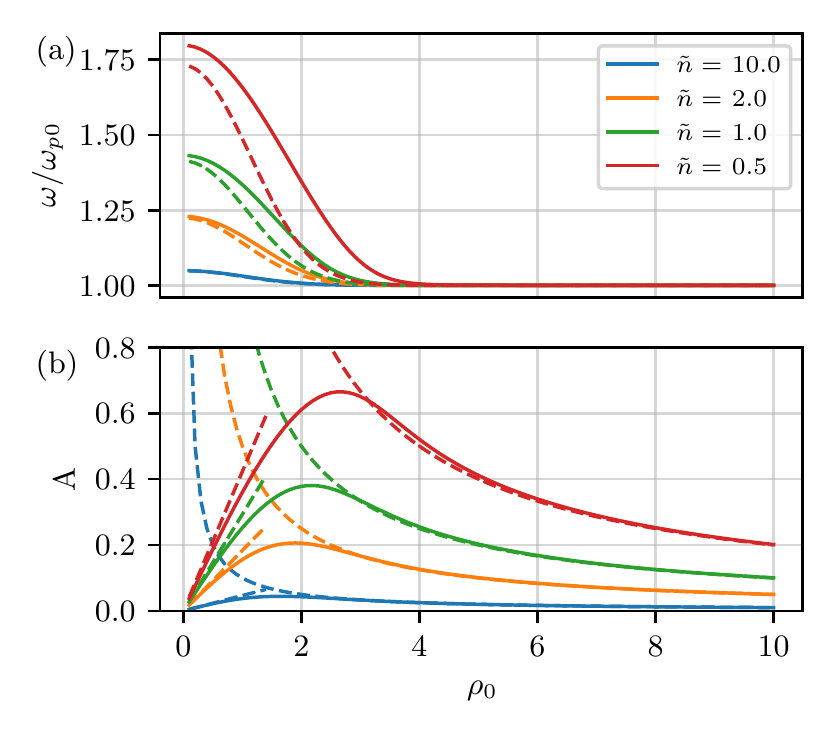}
        \caption{(a) Frequency $\omega$ and (b) amplitude $A$ of plasma electron oscillations as functions of the initial electron radius $\rho_0$ obtained by numerically solving the equation~(\ref{EqMotion3}) for different ratios $n_{b0}/n$. The dashed lines in (a) are approximations (\ref{omega(rho)}) and in (b) are approximations (\ref{A1}) for $\rho_0 \ll 1$ and (\ref{A2}) for $\rho_0 \gg 1$.}\label{fig2}
    \end{figure}

    We consider the following distribution of the driver beam density:
    \begin{equation}\label{BeamDens}
        n_b (\xi, r) = 
        \cases{n_{b0} e^{-r^2/(2 \sigma_r^2)-\xi^2/(2 \sigma_z^2)}, & $\xi<0$, \\
            0, & otherwise.
        }
    \end{equation}
    This beam excites the wave by the steep leading edge that mimics the rapid vapor ionization by a short laser pulse located at $\xi = 0$. Upon entering the plasma section, the proton beam interacts with the radially uniform bounded plasma of radius $R_p \gg \sigma_r$, the density of which varies over a wide range. The interaction regimes gradually change from strongly nonlinear to linear, depending on the density ratio $n_{b0}/n(z)$. To find the trajectories of plasma electrons, we use the one-dimensional Dawson's model for cylindrical nonlinear plasma waves \cite{PR113-383,PoP13-056709}. It implies no crossing of the trajectories, so it is applicable to a limited region in the co-moving frame, where $\xi > \xi_{wb}$. According to the model, the motion of plasma electrons is described in cylindrical coordinates $(r, \xi)$ by the equation
    \begin{equation}\label{EqMotion1}
        \frac{d^{2}r}{d\xi^{2}}=
        \frac{k_{p}^2}{2r}\left[r_{0}^{2}-r^{2}- \int_0^r \frac{n_{b}(\xi, r')}{n} r'dr' \right],
    \end{equation}
    where $r_0$ is the initial radius of the electron at $\xi = 0$, $k_{p} = \sqrt{4\pi n e^2 / (mc^2)}$ is the plasma wavenumber, and $m$ is the electron mass. For the beam density distribution (\ref{BeamDens}), the equation~(\ref{EqMotion1}) takes the form
    \begin{equation}\label{EqMotion2}
        \frac{d^{2}r}{d\xi^{2}}=
        \frac{k_{p}^2}{r}\left[\frac{r_{0}^{2}-r^{2}}{2}-\sigma_{r}^{2} \frac{n_{b0}}{n} \left(1-e^{-\frac{r^{2}}{2 \sigma_{r}^{2}}}\right) e^{-\frac{\xi^2}{2 \sigma_{z}^{2}}}\right].
    \end{equation}
Following the context of the AWAKE, we consider the case of a long proton beam ($\sigma_z \gg k_p^{-1}$). This assumption allows us to neglect the dependence of the beam density on $\xi$ and put $n_b(\xi, r) = n_b(0, r)$ when solving the equation (\ref{EqMotion2}).
    By introducing dimensionless quantities $\rho = r/\sigma_r$, $\rho_0 = r_0/\sigma_r$ and $\tilde{n} = n/n_{b0}$, we simplify the equation (\ref{EqMotion2}) and make it independent of the transverse beam size:
    \begin{equation}\label{EqMotion3}
        \frac{1}{k_p^2} \frac{d^{2}\rho}{d\xi^{2}}=\frac{1}{\rho}\left[\frac{\rho_{0}^{2}-\rho^{2}}{2}-\frac{1}{\tilde{n}} \left(1-e^{-\rho^{2}/2}\right) \right].
    \end{equation}
Solutions of this equation are periodic functions $\rho(\xi)$ which oscillate with frequency $\omega$ and amplitude $A = (\rho_0 - \rho_\mathrm{min})/2$, depending on the initial electron radius~$\rho_0$ (figure~\ref{fig2}), where $\rho_\mathrm{min}$ is the minimum radial position of the electron during oscillations.

    At high plasma densities ($n\gg n_{b0}$), the plasma is able to locally compensate the space charge of the beam. Therefore, the oscillation frequency $\omega$ depends on the radius just as the plasma frequency $\omega_p(n_e) = \sqrt{4\pi n_e e^2 / m}$ calculated for the electron density $n_e = n + n_{b}$:
    \begin{equation}\label{omega(rho)}
        \omega \approx \omega_{p}(n)\sqrt{1 + n_b(0,\rho)/n}
    \end{equation}
    (figure~\ref{fig2}(a)). Compensation of the beam current also changes the oscillation frequency observed in the laboratory frame. This effect is not included into the electrostatic model~(\ref{EqMotion1}). The compensating current flows in a region of radius about $k_p^{-1}$ or $\sigma_r$, whichever is larger \cite{PF13-1831}. If the beam is narrower than the plasma skin depth ($k_p \sigma_r \ll 1$), the current carries the wave pattern as a whole and does not affect the wavebreaking. Otherwise, the current neutralization is local, and the frequency change is twice as large as the equation~(\ref{omega(rho)}) gives.
    At plasma densities $n \lesssim n_{b0}$, the approximation~(\ref{omega(rho)}) is no longer applicable, although it qualitatively agrees with numerical solutions of the equation (\ref{EqMotion3}) (figure~\ref{fig2}(a)). 
    
The amplitude of electron oscillations approximately equals the difference between the initial electron position $\rho_0$ and the equilibrium radius $\rho_\text{eq}$ at which the right-hand side of equation (\ref{EqMotion3}) is zero:
\begin{equation}
    \frac{\rho_{0}^{2}-\rho_\text{eq}^{2}}{2}-\frac{1}{\tilde{n}} \left(1-e^{-\rho_\text{eq}^{2}/2}\right) = 0.
\end{equation}
At small initial radii ($\rho_\text{eq} \sim \rho_0 \ll 1$), we can expand the exponent and obtain
\begin{equation}\label{A1}
    A(\rho_0) \approx \rho_0 - \rho_\text{eq} \approx (1-\sqrt{\tilde{n}/(\tilde{n}+1)})\rho_0.
\end{equation}
At $\rho_\text{eq} \sim \rho_0 \gg 1$, we neglect the exponent and find
\begin{equation}\label{A2}
    A(\rho_0) \approx 1/(\tilde{n}\rho_0),
\end{equation}
so the amplitude $A(\rho_0)$ reaches a maximum in the region $\rho_0 \sim 1$ (figure~\ref{fig2}(b)).

    Differences in the oscillation frequencies of two close plasma electrons with trajectories $\rho_\mathrm{in}(\xi)$ and $\rho_\mathrm{out}(\xi)$, such that $\rho_\mathrm{in}(0) < \rho_\mathrm{out}(0)$, lead to the intersection of the trajectories. When this happens, the radial force driving the initially inner electron rises due to an increase of the total negative electric charge in the region $\rho < \rho_\mathrm{in}(\xi)$. As a consequence, this electron acquires a larger positive radial momentum and changes the character of its motion. All electrons initially located in $\rho_\mathrm{in}(0) < \rho < \rho_\mathrm{out}(0)$ experience similar radial kicks, forming an electron jet that leaves the plasma.

    \begin{figure}[tb]
        \includegraphics[width=\columnwidth]{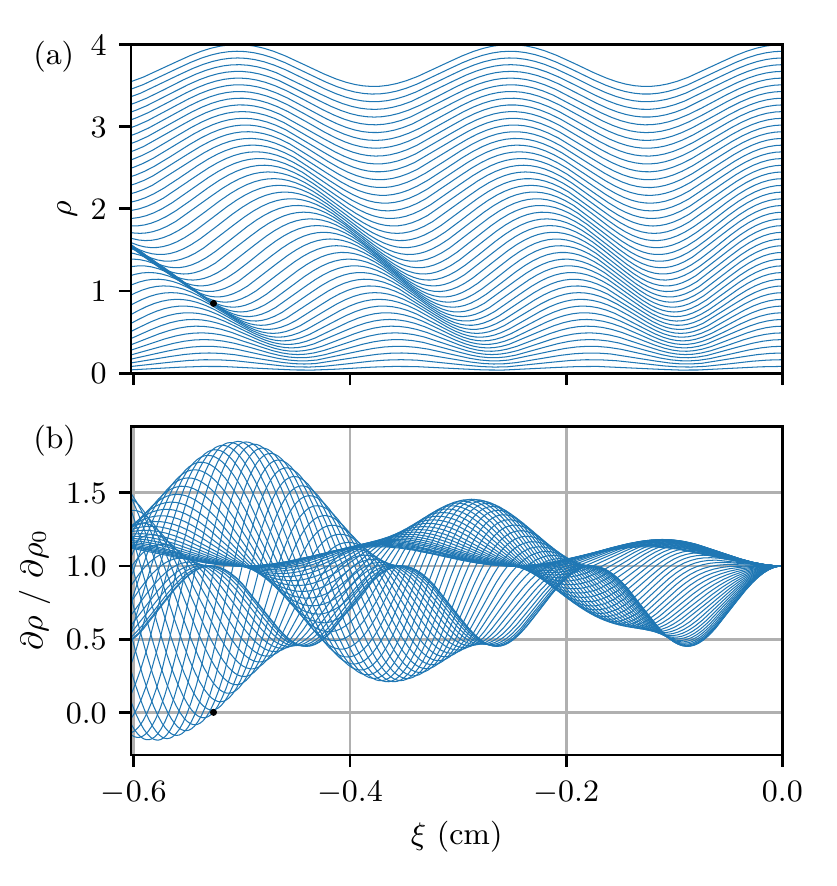}
        \caption{(a) Trajectories of plasma electrons and (b) the corresponding derivatives $\partial\rho/\partial\rho_0$ calculated numerically for \mbox{$\tilde{n} = 1$}. The black dots show the point at which the trajectories intersect for the first time.}\label{fig3}
    \end{figure}    
    The location $\xi_{wb}$ of the point at which the trajectories intersect for the first time depends on the ratio $n_{b0}/n$. To find this point, we differentiate the equation~(\ref{EqMotion3}) by $\rho_0$: 
    \begin{equation}
        \frac{1}{k_p^2} \frac{d^{2}}{d \xi^{2}}\left(\frac{\partial \rho}{\partial \rho_{0}}\right)=\frac{\rho_{0}}{\rho}-\left(\frac{F}{\rho}+\bigl(1+n_{b}(\xi, \rho)/n\bigr)\right) \frac{\partial \rho}{\partial \rho_{0}},\label{r0Deriv}
    \end{equation}
    where $F$ is the right-hand side of the equation~(\ref{EqMotion3}). When neighboring trajectories intersect, the derivative $\partial\rho/\partial\rho_0$ turns to zero. Solving the equation~(\ref{r0Deriv}) numerically (figure~\ref{fig3}(b)), we find the rightmost zero of $\partial \rho/\partial \rho_{0}$ located at $\xi = \xi_{wb}$ (see \cite{notebook} for details). This point defines the trajectories that intersect first. Solving the equation~(\ref{EqMotion3}) for these electrons, we find the radius $r_{wb}$ at which the trajectories intersect (the black dot in figure~\ref{fig3}(a)).

\section{LCODE simulations}
\label{s3}
     
In order to check the accuracy of the Dawson's model and to relate the calculated point ($\xi_{wb}$, $r_{wb}$) to the region where the plasma electron jet is formed, we conducted a series of simulations using a two-dimensional cylindrical quasistatic code LCODE \cite{PRST-AB6-061301,NIMA-829-350}. We put $\sigma_r = 0.2$\,mm, $R_p = 1.1$\,mm and scan the plasma density up to $n_0 = 7\times10^{14}\,\text{cm}^{-3}$, which corresponds to $k_p^{-1} \leq 0.2$\,mm. The simulation grid size is $2\,\mu$m in both $r$ and $\xi$. There are 25 macro-particles per cell for plasma electrons, plasma ions are the immobile background, and the charge and current of the beam are introduced analytically.

    \begin{figure}[tb]
        \includegraphics{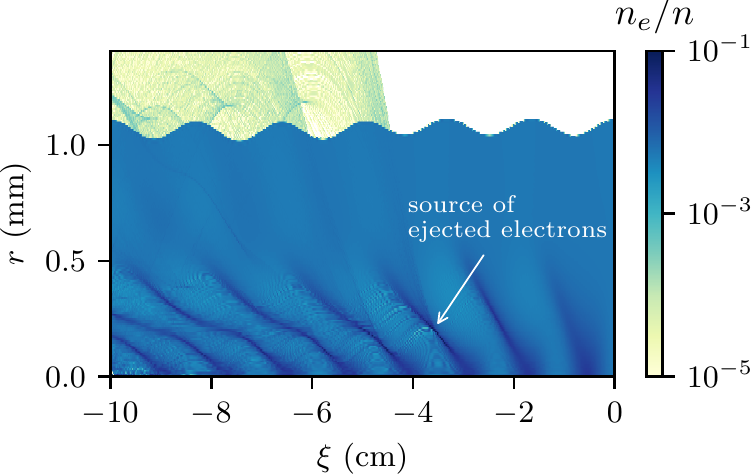}
        \caption{The plasma electron density $n_e (\xi, r)$ obtained from LCODE simulations for $\tilde{n} = 1$. The arrow shows the point in the plasma where the first jet of ejected electrons appears from.}\label{fig4}
    \end{figure}
    \begin{figure}[tb]
        \includegraphics{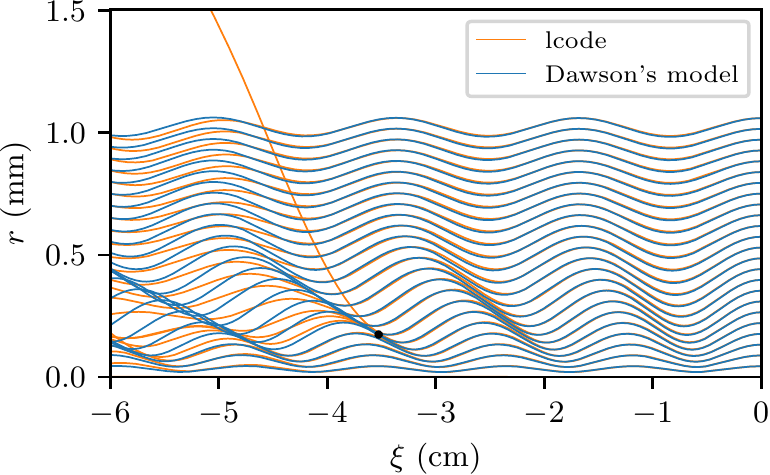}
        \caption{Plasma electron trajectories from LCODE simulations and from the numerical solutions of the equation~(\ref{EqMotion3}) for $\tilde{n} = 1$. The black dot shows the point at which the trajectories intersect for the first time, according to the Dawson's model.}\label{fig5}
    \end{figure}

In typical simulations (figure~\ref{fig4}), jets of ejected electrons appear after several plasma oscillations and are seen both inside and outside the plasma. The point where the first jet originates from perfectly matches the earliest intersection point of the trajectories in the numerical simulation, as well as the results of the Dawson's model (figure~\ref{fig5}). The trajectories calculated using the equation~(\ref{EqMotion3}) fully agree with the simulations, but only at $\xi > \xi_{wb}$, as expected due to the limitations of the model. Thus, our approach makes it possible to localize the point of the first trajectory intersection without using full-scale numerical simulation. 

Plasma electron trajectories that come close to each other cause a local increase in plasma density. Therefore, intersections of neighboring trajectories always occur at high density ridges (figures~\ref{fig4} and~\ref{fig5}). In the case of positively charged beams, the electron trajectories always intersect at points where the radial components of electron velocity are positive. 
Since the motion of plasma electrons is determined by the total linear charge bounded by their trajectories, the intersection changes the charge balance, and particles that had a smaller radius before the intersection experience a weaker return force and continue moving towards the plasma boundary. 

    \begin{figure}[tb]
        \includegraphics{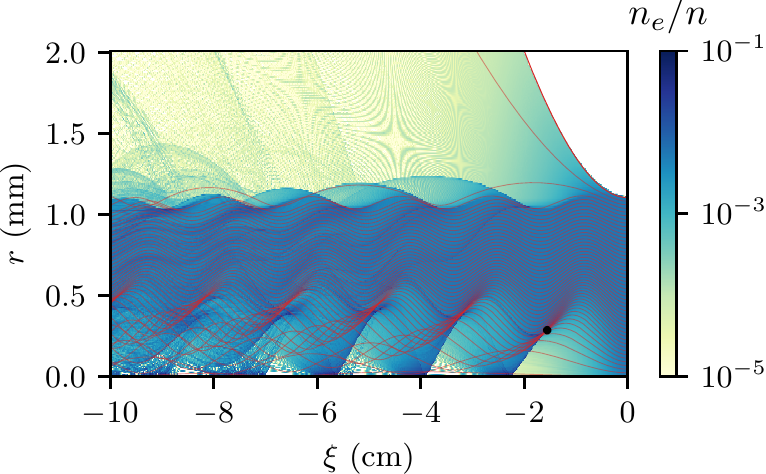}
        \caption{The plasma electron density $n_e (\xi, r)$ obtained from LCODE simulations for the electron driver and $\tilde{n} = 1$. The red lines are solutions of the equation~(\ref{EqMotion3}). The black dot marks the wavebreaking point, according to the Dawson's model.}\label{fig6}
    \end{figure}

For negatively charged beams, the behavior of plasma electrons is qualitatively different (figure~\ref{fig6}). The electrons escape from the plasma immediately after the passage of the beam front.
The total linear charge of the plasma column, including the driver, becomes negative, so the outer electron layer of thickness $\Delta r \approx \sigma_r^2 / (\tilde{n}R)$, estimated for $R \gg \sigma_r$, is ejected to return the plasma to quasineutral state. The motion of the first ejected plasma electron calculated according to the Dawson's model (the upper red curve in figure~\ref{fig6}) fully agrees with the simulations, so the boundary of the electron halo around the plasma can be found semi-analytically.
This process is not related to the intersection of electron trajectories in the plasma wave.
The intersection occurs later and does not necessarily lead to the plasma electron ejection, because the trajectories intersect when the particles move towards the axis.

    \begin{figure}[tb]
        \includegraphics{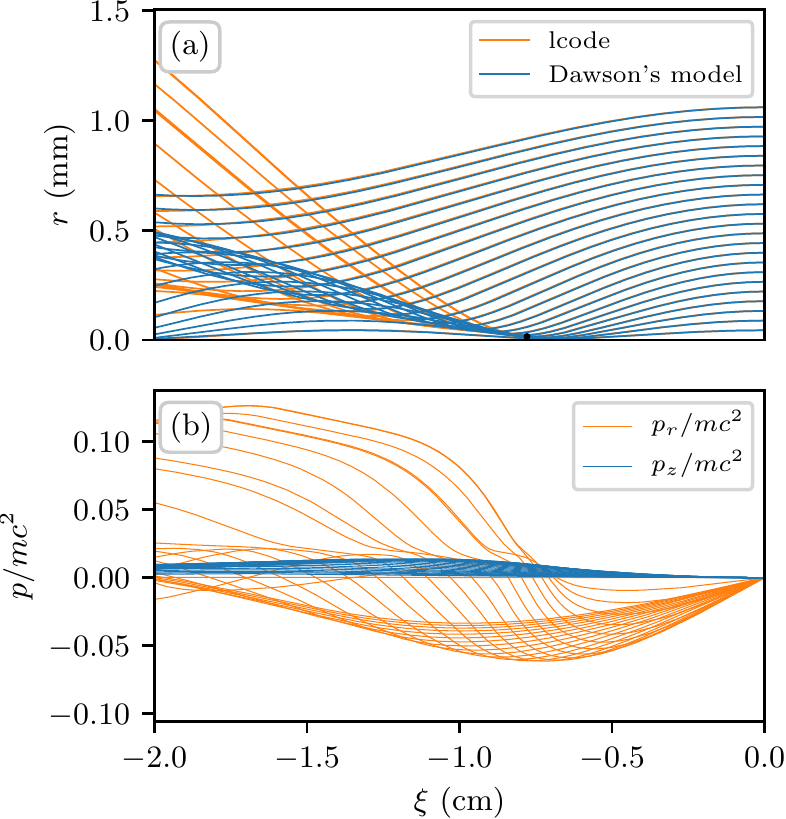}
        \caption{(a) Plasma electron trajectories from LCODE simulations and from the numerical solutions of the equation~(\ref{EqMotion3}) for $\tilde{n} = 0.2$. The black dot shows the point at which the trajectories intersect for the first time, according to the Dawson's model. (b) Components of the electron momentum for these trajectories.}\label{fig7}
    \end{figure}
    \begin{figure}[tb]
        \includegraphics{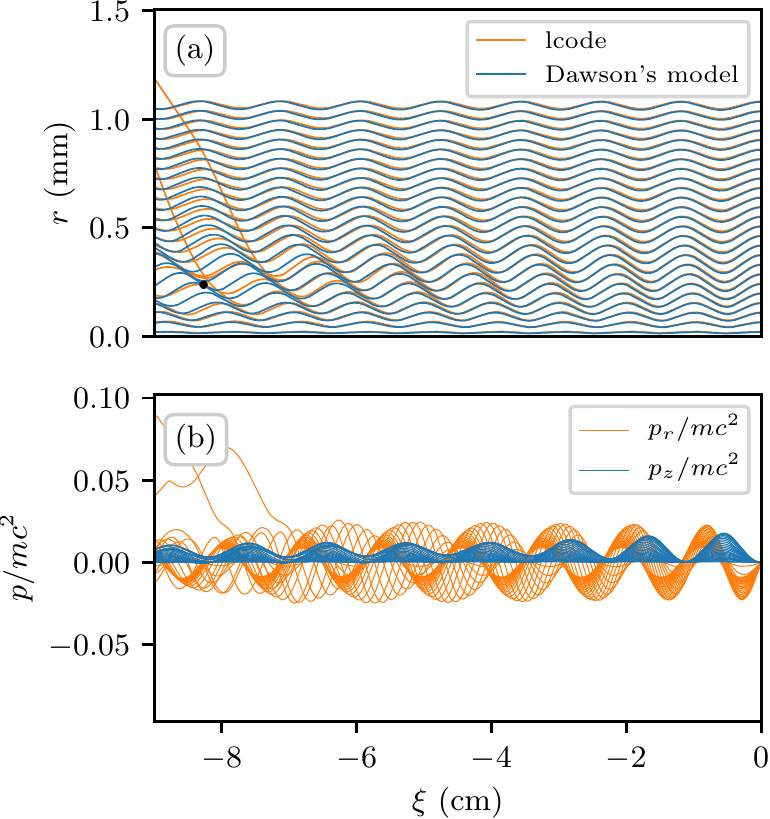}
        \caption{(a) Plasma electron trajectories from LCODE simulations and from the numerical solutions of the equation~(\ref{EqMotion3}) for $\tilde{n} = 2$. The black dot shows the point at which the trajectories intersect for the first time, according to the Dawson's model. (b) Components of the electron momentum for these trajectories.}\label{fig8}
    \end{figure}

The close agreement between the results of the Dawson's model and the numerical simulations shows that the location of the wavebreaking point is mainly determined by the change of the electron oscillation frequency due to the beam charge and, therefore, depends on the ratio of beam and plasma densities. At low plasma densities ($\tilde{n} \lesssim 1$), the plasma frequency on the axis is much higher than at $\rho \sim 1$ (red lines in figure~\ref{fig2}(a)). As a result, the trajectories intersect at the very first period of the plasma wave (figure~\ref{fig7}(a)). The Dawson's model is fully consistent with simulations in this regime. At high plasma densities ($\tilde{n} \gg 1$), the radial variation of the plasma frequency is small (blue lines in figure~\ref{fig2}(a)) and manifests itself only after several plasma periods (figure~\ref{fig8}(a)). However, in simulations, the first intersection of trajectories occurs earlier than in the Dawson's model, which gives only qualitative agreement in this regime. The difference is due to the longitudinal momentum of plasma electrons, which is not taken into account in the equation~(\ref{EqMotion3}). At low plasma densities ($\tilde{n} \ll 1$),  the longitudinal motion of the electrons can be neglected (figure~\ref{fig7}(b)), in contrast to high densities ($\tilde{n} \gg 1$), when the longitudinal and transverse components of the electron velocity are comparable (figure~\ref{fig8}(b)). Note that for the beam parameters considered, the plasma skin depth is equal to the beam radius at $\tilde{n} = 175$. For the variants shown in figures~\ref{fig7} and~\ref{fig8}, $k_p \sigma_r \ll 1$, so the longitudinal velocity affects the wavebreaking by changing the trajectories of oscillating particles, rather than by increasing the frequency difference.

    \begin{figure}[tb]
        \includegraphics{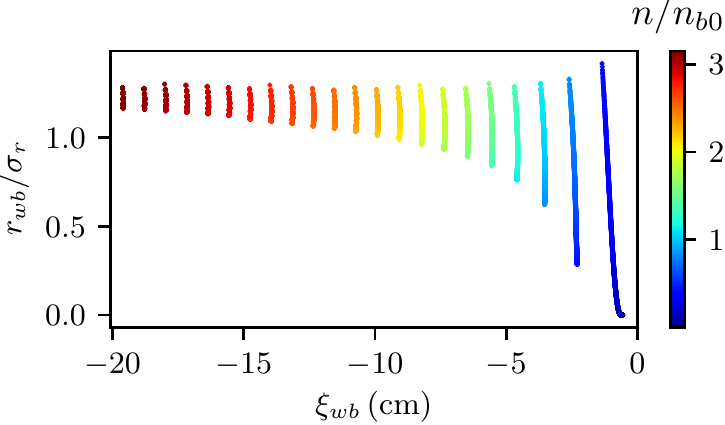}
        \caption{Place of the first intersection of trajectories at different plasma densities, according to the Dawson's model.
}\label{fig9}
    \end{figure}
Despite its inaccuracy at high plasma densities, the Dawson's model allows us to find a qualitative relationship between the position of the first intersection of trajectories and the plasma density (figure~\ref{fig9}). This information is difficult to extract from simulations because the simulated trajectories may intersect due to numerical plasma heating \cite{PPCF63-055002}, making it impossible to algorithmically localize both the longitudinal \cite{PoP25-103103} and the radial coordinates of the first ``real'' wavebreaking event. At low plasma densities ($\tilde{n} \ll 1$), the wave breaks almost on the axis, as in figure~\ref{fig7}(a). As $\tilde{n}$ increases, the point $(\xi_{wb}, r_{wb})$ shifts back and away from the axis, following the shape of the plasma electron density crest. After reaching some threshold radius, which roughly corresponds to the largest beam density gradient, the point jumps to the next density crest, and returns closer to the axis. At $\tilde{n} \gg 1$ the jets of ejected electrons always form at the same radius. 
 
\begin{figure}[tb]
        \includegraphics{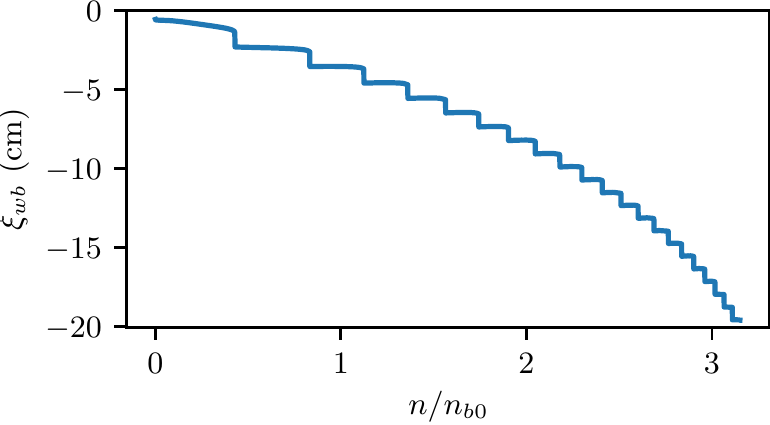}
        \caption{Longitudinal coordinate at which the trajectory crossing occurs for the first time, as a function of the plasma density.}\label{fig10}
    \end{figure}

The dependence of $\xi_{wb}$ on the plasma density $n$ (figure~\ref{fig10}) is of a particular practical importance, since the ejected plasma electrons transfer the wakefields outside the plasma column. Beyond this point (at $\xi < \xi_{wb}$) the wakefields exist outside the plasma column and can distort the trajectory of the injected witness bunch. 
 
 \section{Summary}
A plasma electron halo occurs when a charged particle beam interacts with a radially limited plasma of comparable density. The halo covers a wide area around the plasma column and generates radial forces there. These forces can affect particle beams propagating outside the plasma, deflecting them or degrading their quality. 

In this work, we developed a semi-analytical theory based on the Dawson's sheet model that allows us to clarify the mechanism of halo formation, locate the source of halo electrons, and determine the halo-free region outside the plasma in AWAKE-like setups. The theory agrees with simulation results and shows that, in the case of the proton driver, the halo electrons appear as a result of wavebreaking, that is, the intersection of electron trajectories inside the plasma. For long beams with a sharp leading edge, the location of the wavebreaking point depends only on the transverse size of the beam and the ratio of beam and plasma densities. An increase of the latter postpones the wavebreaking. Negatively charged drivers create the electron halo as soon as they start interacting with plasma. 
At plasma densities much higher then the beam density, the effects of the electron halo become weak for unmodulated beams of any charge sign, since the halo appears later and has a smaller relative charge.
In the future AWAKE experiments, however, the effect of the electron halo will be eliminated by injecting the witness electrons along the axis \cite{JPCS1596-012008} and will not cause issues.

\ack

 This study was funded by RFBR, project number 19-32-90125.

\end{document}